\documentclass[10pt,aps,prc,floatfix,showpacs,twocolumn,longbibliography,superscriptaddress]{revtex4-1}
\usepackage{ulem}
\usepackage{dcolumn}
\usepackage{bm}
\usepackage{color}
\usepackage{amssymb}
\usepackage{amsmath}
\usepackage{graphicx}
\usepackage{amsfonts}
\usepackage{slashed}
\usepackage{pstricks}
\usepackage{hyperref}
\usepackage{array}
\usepackage{multirow}
\usepackage{array}
\allowdisplaybreaks
\usepackage{dcolumn}
\usepackage{epsf}
\usepackage[nice]{nicefrac}
\usepackage{tikz}

\newcommand{\bxi}{\bm \xi}
\newcommand{\br}{\bm r}

\begin{document}
\title{Static and dynamic properties of atomic nuclei with high-resolution potentials }

\author{Alex Gnech}
\affiliation{Department of Physics, Old Dominion University, Norfolk, VA 23529}
\affiliation{Theory Center, Jefferson Lab, Newport News, VA 23610}

\author{Alessandro Lovato}
\affiliation{Physics Division, Argonne National Laboratory, Argonne, IL 60439}
\affiliation{INFN-TIFPA Trento Institute of Fundamental Physics and Applications, 38123 Trento, Italy}

\author{Noemi Rocco}
\affiliation{Theoretical Physics Department, Fermi National Accelerator Laboratory, P.O. Box 500, Batavia, Illinois 60510, USA}

\date{\today}
\begin{abstract}
We compute ground-state and dynamical properties of $^4$He and $^{16}$O nuclei using as input high-resolution, phenomenological nucleon-nucleon and three-nucleon forces that are local in coordinate space. The nuclear Schr\"odinger equation for both nuclei is accurately solved employing the auxiliary-field diffusion Monte Carlo approach. For the $^4$He nucleus, detailed benchmarks are carried out with the hyperspherical harmonics method. In addition to presenting results for the binding energies and radii, we also analyze the momentum distributions of these nuclei and their Euclidean response function corresponding to the isoscalar density transition. The latter quantity is particularly relevant for lepton-nucleus scattering experiments, as it paves the way to quantum Monte Carlo calculations of electroweak response functions of $^{16}$O.
\end{abstract} 
\maketitle

\section{Introduction} 
Tremendous progress has been made in the last two decades in computing properties of atomic nuclei and infinite nuclear matter starting from non-relativistic potentials modeling the individual interactions among protons and neutrons. After the seminal works by Weinberg~\cite{Weinberg:1990rz,Weinberg:1991um}, two- and three-body potentials are nowadays systematically derived starting from an effective Lagrangian, constrained by the broken chiral symmetry pattern of QCD,~\cite{Epelbaum:2008ga,Machleidt:2011zz,Reinert:2017usi,Entem:2017gor}. Chiral effective field theory ($\chi$EFT) potentials serve as input to advanced quantum many-body approaches, suitable to solving the nuclear Schr\"odinger equation~\cite{Carbone:2013eqa,Barrett:2013nh,Hagen:2013nca,Carlson:2014vla,Hergert:2015awm,Lynn:2019rdt}. Among them, methods relying on single-particle basis expansion can treat nuclei up to $^{208}$Pb~\cite{Hu:2021trw,Malbrunot-Ettenauer:2021fnr}, including radii, electromagnetic transitions~\cite{Lechner:2023lyr}, single $\beta$-decay and double-$\beta$ decay rates~\cite{Gysbers:2019uyb}. 

The convergence of the single-particle basis method is greatly aided by the softness of the potential. In addition to the chiral symmetry breaking scale, $\Lambda_{\chi} \approx 1$ GeV, the range of applicability of $\chi$EFT interactions, or their resolution scale, is also determined by a regulator function that smoothly truncates one- and two-pion exchange interactions at short distances. Commonly used chiral forces, such as NNLO${\rm opt}$~\cite{Ekstrom:2013kea}, NNLO${\rm sat}$~\cite{Ekstrom:2015rta}, and $\Delta$NNLO$_{\rm GO}$~\cite{Jiang:2020the}, are characterized by relatively small regulator values. Low-resolution observables like binding energies, radii, and electroweak transitions can be accurately computed using these relatively low-resolution Hamiltonians. In fact, it has been shown that even lower-resolution interactions can reproduce these quantities within a few percent from their experimental values~\cite{Lu:2018bat,Gattobigio:2019omi,  Kievsky:2021ghz,Schiavilla:2021dun,Lovato:2022tjh,Gnech:2023prs}.  

The limitations of $\chi$EFT Hamiltonians become apparent when dealing with observables sensitive to short-range dynamics. An illustrative instance is the equation of state of neutron-star matter at densities beyond nuclear saturation~\cite{Lovato:2022apd}. The Authors of this reference found that certain local, $\Delta$-full $\chi$EFT Hamiltonians, that accurately reproduce the light nuclei spectrum~\cite{Piarulli:2017dwd,Baroni:2018fdn}, yield self-bound neutron matter~\cite{Lovato:2022apd}. This un-physical behavior, characterized by closely packed neutron clusters, arises from the attractive nature of the three-body force contact term. The latter does not vanish in neutron matter because of regulator artifacts~\cite{Lovato:2011ij}. Similarly, a dramatic overbinding of nuclear matter and no saturation at reasonable densities is obtained using the set of $\chi$EFT potentials~\cite{Huther:2019ont} that reproduce accurately experimental energies and radii of nuclei up to $^{78}$Ni~\cite{Sammarruca:2020jsp}.

Highly realistic, phenomenological nucleon-nucleon (NN) forces, such as the Argonne $v_{18}$ (AV18)~\cite{Wiringa:1994wb} are constructed to reproduce nucleon-nucleon scattering data with $\chi^2\sim 1$ up to energies corresponding to pion-production threshold. However, their predictions remain accurate even at larger energies~\cite{Benhar:2021doc}, so as they can be identified as ``high-resolution'' potentials. The remarkable accuracy of the Argonne interactions has been recently highlighted by the analysis of semi-exclusive electron-scattering data~\cite{CLAS:2020mom}, which are extremely sensitive to short-range correlated pairs in nuclei. In addition, their connection with the equation of state of neutron-star matter has been pointed out in Ref.~\cite{Benhar:2021doc}. Nonetheless, these phenomenological potentials have several shortcomings. The Argonne $v_{18}$ plus Illinois 7~\cite{Pieper:2008rui} (IL7) Hamiltonian can reproduce the experimental energies of nuclei with up to $A=12$ with high precision, but it fails to provide sufficient repulsion in pure neutron matter~\cite{Maris:2013rgq}. On the other hand, the Argonne $v_{18}$ plus Urbana IX (UIX) model, while providing a good description of nuclear-matter properties, does not satisfactorily reproduce the spectrum of light nuclei~\cite{Pieper:2001ap}. As these Hamiltonians are phenomenological in nature, no clear prescriptions are available to properly assess their uncertainties and to systematically improve them.

In this work, we carry out high-resolution calculation of $^4$He, and $^{16}$O static and dynamic properties, including binding energies, radii, momentum distribution and isoscalar density response functions. We focus on phenomenological Hamiltonians of the Argonne family, specifically on the Argonne $v_8^\prime$ (AV8P) and Argonne $v_6^\prime$ (AV6P) NN forces~\cite{Wiringa:2002ja} supplemented by the consistent Urbana IX (UIX) 3N interaction~\cite{Pudliner:1995wk}. To gauge the accuracy in solving the quantum many-body problem, for the $^4$He nucleus, we benchmark the auxiliary-field diffusion Monte Carlo (AFDMC) with the highly-accurate hyper-spherical harmonics (HH) method~\cite{Kievsky:2008es,Marcucci:2019hml}. This comparison is particularly relevant for quantifying the approximations made in the imaginary-time propagation upon which the AFDMC is based. 
We note that useful benchmarks among different many-body methods have been carried out for both nuclei~\cite{Kamada:2001tv,Hagen:2007ew,Tichai:2020dna}, and infinite neutron matter Ref.~\cite{Baldo:2012nh,Piarulli:2019pfq, Lovato:2022apd}.

This article is organized as follows. In Section~\ref{sec:methods} we introduce the nuclear Hamiltonian and the quantum many-body methods of choice. Section ~\ref{sec:results} is devoted to the results, including a detailed comparison between the HH and AFDMC methods. Finally, in Section~\ref{sec:conclusions} we draw our conclusions and discuss future perspectives of this work.

\section{Methods}
\label{sec:methods}

\subsection{Phenomenological Nuclear Hamiltonians}
To a remarkably large extent, the dynamics of atomic nuclei can be accurately modeled employing a non-relativistic Hamiltonian 
\begin{align}
H = \sum_i \frac{{\bf p}_i^2}{2m} + \sum_{i<j}^{A} v_{ij}
 + \sum_{i<j<k} V_{ijk} \ ,
\label{H:A}
\end{align}
where ${\bf p}_i$ and $m$ denote the momentum of the $i$-th nucleon and its
mass, and the potentials $v_{ij}$ and $V_{ijk}$ describe nucleon-nucleon (NN) and three-nucleon (3N) interactions, respectively. Highly-realistic, phenomenological NN potentials, such as the AV18~\cite{Wiringa:1994wb}, are fitted to reproduce the observed properties of the
two-nucleon system, including the deuteron binding energy, magnetic moment and electric quadrupole moment, as well as the data obtained from the measured NN scattering cross sections\textemdash and reduce to Yukawa's one-pion-exchange potential at large distance. They are usually defined in coordinate space as
\begin{equation}
v_{ij} = \sum_{p=1}^{18} v^p(r_{ij})O^p_{ij}
\label{eq:vNN}
\end{equation}
with $r_{ij}= |{\bf r}_i - {\bf r}_j|$.
The bulk of the $N\!N$ interaction is encoded in the first eight operators
\begin{equation}
O^{p=1-8}_{ij}= [1, \sigma_{ij},S_{ij},\mathbf{L}\cdot\mathbf{S}]\otimes [1,\tau_{ij} ] \, .
\label{eq:oper}
\end{equation}%
In the above equation we introduced $\sigma_{ij}={\bm \sigma_i}\cdot{\bm \sigma_j}$ and $\tau_{ij}={\bm \tau_i}\cdot{\bm \tau_j}$ with ${\bm \sigma}_i$ and ${\bm \tau}_i$ being the Pauli matrices acting in the spin and isospin space. The tensor operator is given by
\begin{equation}
S_{ij}= \frac{3}{r^2_{ij}}({\bm \sigma}_i\cdot {\bf r}_{ij})({\bm \sigma}_j \cdot{\bf r}_{ij})- \sigma_{ij}\, ,
\end{equation}
while the spin-orbit contribution is expressed in terms of the relative angular momentum $\mathbf{L}=\frac{1}{2 i} (\mathbf{r}_i -\mathbf{r}_j) \times (\nabla_i - \nabla_j)$ and the total spin $\mathbf{S}=\frac{1}{2}({\bm \sigma}_i+{\bm \sigma}_j)$ of the pair. AV18 contains six additional charge-independent operators corresponding to $p=9-14$ that are quadratic in $\mathbf{L}$, while the $p=15-18$ are charge-independence breaking terms.

It is useful to define simpler versions of the AV18 potentials with 
fewer operators: the $v^\prime_8$ (AV8P) with the eight operators of Eq.~\eqref{eq:oper}
and the $v^\prime_6$ (AV6P) without the $\mathbf{L}\cdot\mathbf{S}\otimes [1,\tau_{ij} ]$
terms~\cite{Wiringa:2002ja}.  
AV8P is a reprojection (rather than a simple truncation)
of the strong-interaction potential that reproduces the charge-independent
average of $^1$S$_0$, $^3$S$_1$-$^3$D$_1$, $^1$P$_1$, $^3$P$_0$, $^3$P$_1$, 
and (almost) $^3$P$_2$ phase shifts by construction, while overbinding the
deuteron by $18$ keV due to the omission of electromagnetic terms.
AV6P is (mostly) a truncation of AV8P that reproduces
$^1$S$_0$ and $^1$P$_1$ partial waves, makes a slight adjustment to (almost)
match the $v^\prime_8$ deuteron and $^3$S$_1$-$^3$D$_1$ partial waves, but will no longer split the $^3$P$_J$ partial waves properly.

\begin{figure*}[htb]
\includegraphics[width=\textwidth]{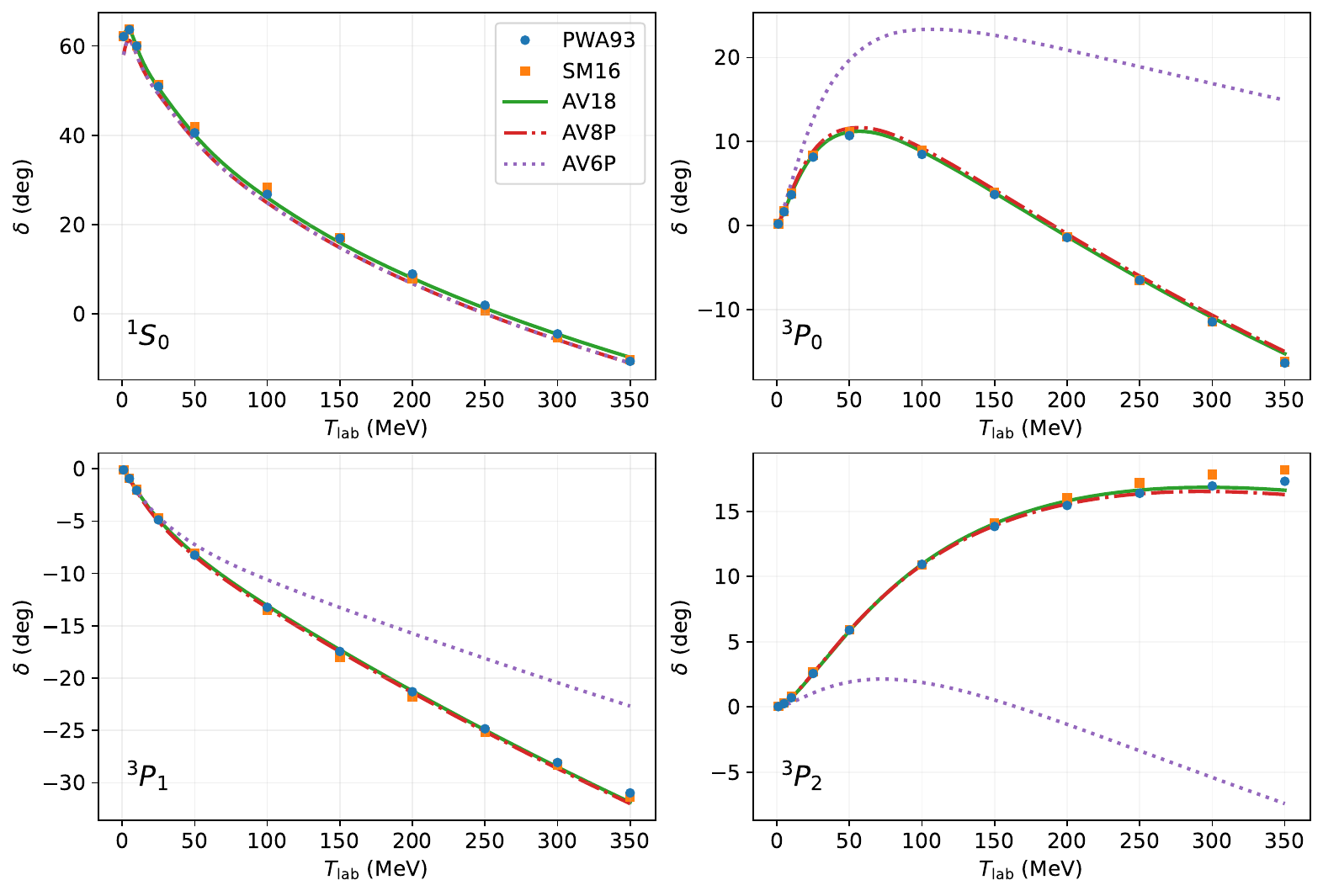}
\caption{Neutron-proton scattering phase shifts in the $^1$S$_0$, $^3$P$_0$, $^3$P$_1$, and $^3$P$_2$ channels are depicted as a function of the kinetic energy of the beam particle in the laboratory frame. The long-dashed-dotted, solid, dot-dashed, and dashed lines represent the AV18, AV8$^\prime$, and AV6$^\prime$ predictions. The solid dots and squares are derived from the PWA93~\cite{Stoks:1993tb} and SM16~\cite{Workman:2016ysf} analyses, respectively.}
\label{fig:phase_shifts}
\end{figure*}

Fig.\ref{fig:phase_shifts} illustrates the energy dependence of the proton-neutron scattering phase shifts in the $^1$S$_0$, $^3$P$_0$, $^3$P$_1$, and $^3$P$_2$ partial waves, comparing the AV6P, AV8P, and AV18 potentials with the experimental analysis of Refs.\cite{Stoks:1993tb,Workman:2016ysf}. The AV18 interaction provides an accurate description of the scattering data up to $T_{\rm lab} = 350$ MeV in all channels. All potentials reproduce the $^1S_0$ channel very well, as neither spin-orbit nor tensor components are active there. On the other hand, in the $^3P_0$, $^3P_1$, and $^3P_2$ channels, the AV6P potential deviates from the AV18 results, particularly in the $^3P_2$ wave. These shortcomings, which can be ascribed to the missing spin-orbit components, are not present in the AV8P potential, which reproduces all $P$-wave phase shifts. Discrepancies between AV8P and AV18 appear in higher partial waves. 

In this work,  we assume the electromagnetic component of the NN potential to only include the Coulomb force between finite-size protons. 

The earliest example of a three-body force in nuclear physics dates back to the Fujita-Miyazawa interaction, whose main contributions arises from the virtual excitation of a $\Delta(1232)$ resonance in processes involving three interacting nucleons~\cite{Fujita:1957zz}. This term is included in the UIX potential as 
\begin{equation}
V_{ijk}^\Delta = V_{ijk}^{\Delta,a} + V_{ijk}^{\Delta,c}\, .
\end{equation}
The ``commutator'' and ``commutator'' contributions are given by
\begin{align}
V_{ijk}^{\Delta,a} &= \sum_{\rm cyc} A_{2\pi}\{X_{ij}^\pi,X_{jk}^\pi\}\{\tau_{ij},\tau_{jk}\} \nonumber\\
V_{ijk}^{\Delta,c} &= \sum_{\rm cyc} C_{2\pi}[X_{ij}^\pi,X_{jk}^\pi][\tau_{ij},\tau_{jk}] \,.,
\end{align}
where $\sum_{\rm cyc}$ denotes a sum over the three cyclic exchanges of nucleons $i,j,k$.  The operator ${X}^\pi_{ij}$ is defined as 
\begin{equation}
X_{ij}^\pi = T(r_{ij})\, S_{ij} + Y(r_{ij})\sigma_{ij} \, ,
\end{equation}
where the normal Yukawa and tensor functions are
\begin{align}
Y(r)&=\frac{e^{-{m_{\pi}}r}}{{m_{\pi}}r} \,\xi(r) \nonumber\\
T(r)&=\left( 1 + \frac{3}{m_{\pi}\,r} + \frac{3}{m_{\pi}^2\, r^2} \right)  Y(r) \, .
\end{align}
and the short-range regulator is taken to be
$\xi(r) = 1 - \exp(-c r^2)$ with $c =2.1$ fm$^{-2}$. In the UIX model, it is assumed that $C_{2\pi} = A_{2\pi} / 4$ as in the original Fujita-Miyazawa formulation. We note however that other ratios slightly larger than $1/4$ have been considered in the Tucson-Melbourne~\cite{Coon:1978gr} and Brazil~\cite{Coelho:1983woa} models of the 3N force. Generally, a 3N potential only comprised of $V_{ijk}^\Delta$ does not yield the correct isospin-symmetric nucleonic matter saturation density~\cite{Lagaris:1981mn} and a repulsive 3N force should be included 
\begin{equation}
V^{R}_{ijk} = A_R \sum_{\rm cyc} T^2(r_{ij}) T^2(r_{jk}) \,.
\end{equation}
The constants $A_{2\pi}$ and $A_R$ entering the UIX model are determined by fitting the binding energy of $^3$H and the saturation density of isospin-symmetric nucleonic matter $\rho_0 = 0.16$ fm$^{-3}$. More sophisticated phenomenological 3N interactions have been developed, such as the IL7~\cite{Pieper:2008rui}, but we will not consider them here as they fail to provide sufficient repulsion in pure neutron matter~\cite{Maris:2013rgq}. Rather, we stick to the UIX model, although it underbinds light p-shell nuclei --- and in particular it is not possible to fit both $^8$He and $^8$Be at the same time~\cite{Carlson:2014vla}.

\subsection{Auxiliary field diffusion Monte Carlo}
The AFDMC method~\cite{Schmidt:1999lik} leverages imaginary-time projection techniques to filter out the ground-state of the system starting from a suitable variational wave function
\begin{equation}
|\Psi_0\rangle = \lim_{\tau \to \infty} |\Psi(\tau)\rangle = \lim_{\tau \to \infty} e^{-(H-E_V)\tau} |\Psi_V\rangle\, .
\end{equation}
In the above equation $E_V$ is a normalization constant, which is chosen to be close to the true ground-state energy $E_0$. 

The variational wave function is expressed as a product of a long and a short-range parts $|\Psi_V\rangle = \hat{F} |\Phi\rangle$. To fix the notation, let $X=\{x_1\dots x_A\}$ denote the set of single-particle coordinates $x_i = \{\mathbf{r}_i, \sigma_i, \tau_i\}$, which describe the spatial positions and the spin-isospin degrees of freedom of the $A$ nucleons. The long-range behavior of the wave function is described by the Slater determinant
\begin{equation}
\langle X |\Phi\rangle=\mathcal{A}\{\phi_{\alpha_1}(x_1),\dots,\phi_{\alpha_A}(x_A)\}
\, .
\end{equation}
The symbol $\mathcal{A}$ denotes the antisymmetrization operator and 
$\alpha$ represent the quantum numbers of the 
single-particle orbitals. The latter are expressed as
\begin{equation}
\phi_{\alpha}(x)=R_{nj}(r) \, [Y_{\ell m_\ell}(\hat{r})\, 
\, \chi_{s m_s}(\sigma)]_{jm_j}\, \chi_{t m_t}(\tau)
\, ,
\end{equation}
where the spherical harmonics $Y_{\ell \ell_z}(\hat{r})$ are coupled to the spinor $\chi_{ss_z}(\sigma)$  to get the single-particle orbitals in the $j$ basis and $\chi_{\tau \tau_z}(\tau)$ describes the isospin single-particle state. The radial functions $R_{nj}(r)$ are parametrized in terms of cubic splines~\cite{Contessi:2017rww}. 

Since the Hamiltonians considered in this work contain tensor and spin-orbit interactions, we consider a variational wave function that includes two- and three-body correlations 
\begin{align}
\hat{F} &= \prod_{i<j}f^{2b}(r_{ij}) \prod_{i<j<k} f^{3b}_{ijk} \nonumber\\
&\times \Big[1 + \sum_{i<j}U^{2b}_{ij} + \sum_{i<j<k}U^{3b}_{ijk}\Big]
\label{eq:corr}
\end{align}
To keep a polynomial cost in the number of nucleons, as in Ref.~\cite{Gandolfi:2014ewa}, we approximate the spin-isospin correlations to a linear form
\begin{equation}
U^{2b}_{ij} = 1 + \sum_p u^p_{\rm 2b}(r_{ij})O^p_{ij}\,.
\end{equation}
where the operators $O^p_{ij}$ are defined in Eq.~\eqref{eq:oper}. The functions $u^p(r)$ are characterized by a number of variational parameters~\cite{Gandolfi:2020pbj}, which are determined by minimizing the two-body cluster contribution to the energy per particle of nuclear matter at saturation density.

The spin-isospin dependent three-body correlations are derived within perturbation theory as~\cite{Carlson:2014vla, Lonardoni:2018nob, Gandolfi:2020pbj} 
\begin{equation}
U^{3b}_{ijk} = - \epsilon V^{\Delta, a}_{ijk}\, ,
\end{equation}
with $\epsilon$ being a variational parameter. Finally, the scalar three-body correlation is expressed as 
\begin{equation}
f^{3b}_{ijk} = 1 - \sum_{cyc} u_{\rm 3b}(r_{ij})u_{\rm 3b}(r_{jk})\,.
\end{equation}
Both the three-body correlations $u_{\rm 3b}(r)$, the scalar two-body Jastrow $f^{\rm 2b}(r)$ are parametrized in terms of cubic splines~\cite{Contessi:2017rww}. The variational parameters are the values of $u_{\rm 3b}(r)$ and $f^{\rm 2b}(r)$ at the grid points, plus the value of their first derivatives at $r = 0$. The optimal values of the variational parameters are found employing the linear optimization method~\cite{Toulouse:2007,Contessi:2017rww}, which typically converges in about $20$ iterations. We note that variational states based on neural network quantum states have been recently proposed~\cite{Adams:2020aax,Gnech:2021wfn,Lovato:2022tjh,Gnech:2023prs}. Their use in AFDMC calculations will be the subject of future works. 

The imaginary-time propagator $e^{-(H-E_V)\tau}$ is broken down in $N$ small time steps $\delta \tau$, with $\tau= N \delta\tau$. At each step, the generalized coordinates $X^\prime$ are sampled from the previous ones according to the short-time propagator  
\begin{equation}
G(X^\prime,X,\delta\tau) = \frac{\Psi_I(X^\prime)}{\Psi_I(X)} \langle X^\prime| e^{-(H-E_0)\delta\tau}|  X \rangle
\end{equation}
where $\Psi_I(X^\prime)$ is the importance-sampling function. Similar to Refs.~\cite{Zhang:1996us,Zhang:2003zzk}, we mitigate the fermion-sign problem by first performing a constrained-path diffusion Monte Carlo propagation (DMC-CP), in which we take $\Psi_I(X) \equiv \Psi_T(X)$ and impose ${\rm Re}[\Psi_T(X^\prime)/\Psi_T(X)] > 0$. 
The energy computed during a DMC-CP is not a rigorous upper-bound to $E_0$~\cite{Wiringa:2000gb}. To remove this bias, we further evolve the DMC-CP configuration using the positive-definite importance sampling function~\cite{Pederiva:2004iz,Lonardoni:2018nob,Piarulli:2019pfq}
\begin{align}
\Psi_I(X) = \sqrt{ \rm Re[\Psi_T(X)]^2 + \rm Im [\Psi_T(X)]^2}\,,
\end{align}
During this unconstrained diffusion (DMC-UC), the asymptotic value of the energy is determined by fitting its imaginary-time behavior with a single-exponential function as in Ref.~\cite{Pudliner:1997ck}. 

The AFDMC keeps the computational cost polynomial in the number of nucleons $A$ by representing the spin-isospin degrees of freedom in terms of outer products of single-particle states. To preserve this representation, during the imaginary-time propagation Hubbard-Stratonovich transformations are employed to linearize  the quadratic spin-isospin operators entering realistic nuclear potentials. While the AV6P interaction can be treated exactly, applying these transformation to treat the isospin-dependent spin-orbit term entering the AV8P potential involves non-trivial difficulties. To circumvent them, we perform the imaginary-time propagation with a modified NN interaction in which
\begin{equation}
v^8(r_{ij}) \mathbf{L}\cdot\mathbf{S} \times \tau_{ij} \to 
\alpha v^8(r_{ij}) \mathbf{L}\cdot\mathbf{S}\,.
\label{eq:spin_orbit_prop}
\end{equation}
The parameter $\alpha$ is adjusted by making the expectation value of the original and modified AV8P interaction the same $\langle v^8(r_{ij}) \mathbf{L}\cdot\mathbf{S} \times \tau_{ij} \rangle = \alpha \langle v^8(r_{ij}) \mathbf{L}\cdot\mathbf{S} \rangle$. 

The commutator term of $V_{ijk}^\Delta$ entails cubic spin-isospin operators, which presently cannot be encompassed by continuous Hubbard-Stratonovich transformations. Restricted Boltzmann machines offer a possible solution to this problem~\cite{Rrapaj:2020txq} and their application to AFDMC calculations is currently underway and will be the subject of a future work. Here, we follow a similar strategy as for the isospin-dependent spin-orbit term of the NN potential and perform the imaginary-time propagation with a 3N potential modified as
\begin{equation}
V_{ijk}^{\Delta,c } \to \beta  V_{ijk}^{\Delta,a}
\label{eq:cubic_prop}
\end{equation}
Again, $\beta$ is adjusted such that the expectation value of the original and modified $V_{ijk}^\Delta$ be the same $\langle V_{ijk}^{\Delta,c }  \rangle = \beta \langle V_{ijk}^{\Delta,a }  \rangle$.

\subsection{Hyperspherical harmonics}
In this work we use the hyperspherical harmonic (HH) method developed by the Pisa group for $A=4$~\cite{Kievsky:2008es,Marcucci:2019hml}. 
In the HH method the $A=4$ Hamiltonian is rewritten using the Jacobi coordinates. This permits to
completely decouple the center of mass motion. A suitable choice for the Jacobi coordinates is
\begin{align}
    \bxi_{1p}&=\sqrt{\frac{3}{2}}\left(\br_l-\frac{\br_i+\br_j+\br_k}{3}\right)\\
    \bxi_{2p}&=\sqrt{\frac{4}{3}}\left(\br_k-\frac{\br_i+\br_j}{2}\right)\\
    \bxi_{3p}&=\left(\br_i-\br_j\right)
\end{align}
where $p$ specify a given permutation of the particles $i,j,k,l$.
The kinetic energy opertor is then rewritten in terms of the hyperpherical coordinates given by  the hyperradius $\rho=\sqrt{\bxi_{1p}^2+\bxi_{2p}^2+\bxi_{3p}^2}$, and the hyperangular coordinates $\Omega_N^{(p)}=\{\hat \xi_{1p},\hat \xi_{2p},\hat \xi_{3p},\phi_{1p},\phi_{2p}\}$ where $\hat\xi_{ip}$ are the polar angles of the Jacobi coordinates, and  $\cos \phi_{i}=\frac{\xi_{i}}{\sqrt{\xi_{1}^2+\dots+\xi_{i}^2}}$. In this way, it is possible to isolate in the kinetic energy an operator $\Lambda^2(\Omega_N^{(p)})$ that depends only on the hyperangular coordinates taht is  known as  the grandangular momentum operator. The eigenstates of this operator are the hyperspherical functions 
\begin{equation}
    {\cal Y}_{[K]}=\left[\left( Y_{\ell_1}(\xi_{1p})Y_{\ell_2}(\xi_{2p})\right)_{L_2}Y_{\ell_3}(\xi_{3p})\right]_{L,L_z}{\cal P}^{n_1,n_2}_{\ell_1,\ell_2,\ell_3}(\phi_1,\phi_2)
\end{equation}
where $Y_{\ell_i}$ are the standard spherical harmonics and ${\cal P}^{n_1,n_2}_{\ell_1,\ell_2,\ell_3}(\phi_1,\phi_2)$ is a specific normalized combination of Jacobi polynomials (see Refs.~\cite{Kievsky:2008es,Marcucci:2019hml} for the full expression).
Note that we construct the hyperspherical functions such that they are the eigenstate of the total angular momentum  operator with eigenvalues $L$ and $L_z$ respectively. We use the symbol $[K]=\{\ell_1,\ell_2,\ell_3,L_2,L,L_z,n_2,n_3\}$ to represent the full set of quantum numbers needed to uniquely identify the hyperspherical state. The value $K=\ell_1+\ell_2+\ell_3+2n_1+2n_2$ is the eigenvalue of the operator $\Lambda^2(\Omega_N^{(p)})$ associated with the eigenstate $ {\cal Y}_{[K]}$.

The $^4$He wave function is then written as 
\begin{equation}
\Psi_4=\sum_{l,\alpha}c_{l,\alpha}f_l(\rho)\sum_{p=1}^{12}Y^{JJ_z,TT_z}_{\alpha}(\Omega_N^{(p)})\,.
\label{eq:psihh}
  \end{equation}
where
     \begin{equation}
Y^{JJ_z,TT_z\pi}_{[\alpha]}=\left[{\cal Y}_{[K]}(\Omega_N^{(p)})\chi^{p}_{[S]}\right]_{JJ_z}\chi^p_{[T]}\,,
    \label{eq:hhst-basis}
  \end{equation}
is the HH+spin+isospin basis with fixed total angular momentum $J,J_z$, parity $\pi$ and total isospin $T,T_z$.
The spin part is given by
\begin{equation}
    \chi^{p}_{[S]}=\left[\left[(s_i s_j)_{S_2} s_k\right]_{S_3}s_l\right]_{S,S_z}\,,
\end{equation}
where $s_i$ is the spin of the single particle, and  with the symbol $[S]=\{S_2,S_3,S,S_z\}$ we indicate the full set of quantum numbers needed to uniquely identify the spin state. The isospin state is written in an analogous way. We use the symbol $\alpha=\{[K],[S],[T]\}$ to indicate the full set of quantum numbers we use to describe an element of the HH+spin+isospin basis. The sum over the 12 even permutation $p$ is then exploited to fully antisymmetrize the basis set (see Refs.~\cite{Kievsky:2008es,Marcucci:2019hml}).

The expansion on the hyperradial part is performed using the function $f_l(\rho)$ that in our case reads
\begin{equation}
    f_l(\rho)=\gamma^{\frac{9}{2}}\sqrt{\frac{l!}{(l+8)!}}L_l^{(8)}(\gamma\rho){\text e}^{-\gamma\rho/2}\,,
\end{equation}
where $L_l^{(8)}$ are the generalized Laguerre polynomials and $\gamma$ is a non-variational parameter which value is typically between $3-5$ fm$^{-1}$.
Finally, the coefficients $c_{l,\alpha}$ are unknown and they are determined solving the eigenvalue-eigenvector problem derived from the Rayleigh-Ritz variational principle.
 
The use of the Argonne family potential for the $A=4$ system is quite challenging for the HH method. Due to strong short range repulsion of the potential, a large value of the grandangular momentum $K$ is required to reach convergence~\cite{Viviani:2004vf}. This make impossible to include in our basis all the HH states up to given value of $K$. In this work we follow the approach used in Ref.~\cite{Gnech:2020qtt}, where an effective subset of the HH basis is selected, based on two elements: i) the importance of the partial waves appearing in the $^4$He wave function (i.e the quantum numbers $L,S,T$), and ii) the total centrifugal barrier of the HH states $\ell_{tot}=\ell_1+\ell_2+\ell_3$. In Table~\ref{tab:HH_class} we report the maximum value of $K$ used for each class of HH basis defined by the quantum numbers $L,S,T,\ell_{tot}$.

\begin{table}[!htb]
    \centering
    \begin{tabular}{l|cccc}
    \hline
    \hline
     $L,S,T$ & $\ell_{tot}=0$ & $\ell_{tot}=2$ & $\ell_{tot}=4$ & $\ell_{tot}\geq 6$ \\
     \hline
      0,0,0 & 48 & 34 & 28 & 20 \\
      2,2,0 & -- & 34(48 for $\ell_1+\ell_2=0$) & 28 & 20 \\
      1,1,0 & -- & 34 & 28 & 20 \\
      \hline
      \hline
    \end{tabular}
    \caption{Maximum value of $K$ used for each class subset of the HH basis defined by the quantum numbers $L,S,T,\ell_{tot}$}
    \label{tab:HH_class}
\end{table}
Using this basis subset with both the AV6P and the AV8P we are able to reach convergence below 10 KeV on the binding energies when only  the two body forces are used. In Table~\ref{tab:energies_4He} we maintain a conservative 10 KeV error.

Let us now consider the three-body forces.
Since the radial part of the three-body forces is rather soft, the correlations induced by the 3N interaction are not generating large grand angular quantum number components~\cite{Viviani:2004vf}.  Therefore, in our calculation we included the three-body forces up to $K_{3N}=18$ neglecting larger components. In this case, we were able to reach an accuracy on the binding energy of the level of 50 KeV. Studying the convergence pattern of the binding energy using different values of $K_{3N}$ for the cut on the three body forces and maintaining the two-body part fully converged we extrapolate the binding energy for $K_{3N}\rightarrow\infty$ . Our final extrapolation with the associated error is reported in Table~\ref{tab:energies_4He}.

\section{Results and Discussion}
\label{sec:results} 
\subsection{Ground-state energies}

\begin{table}[b]
\renewcommand{\arraystretch}{1.25} 
\centering
\begin{tabular}{ c | c c c c c }
\hline
\hline
Hamiltonian & HH  &  VMC & DMC-CP &  DMC-UC\\
\hline
AV6P & -26.13(1)  & -22.87(5)  & -26.37(6) & -26.17(10)\\
AV8P     & -25.15(1)  &   -21.50(8)  & -25.0(4) & -24.5(6)\\
AV6P+UIX & -31.45(1)  &   -25.64(6)  & -31.6(6) & -31.4(7)\\
AV8P+UIX & -29.92(1)  &   -24.16(7)  & -29.9(8) & -29.4(9) \\
\hline
\end{tabular}
\caption{Ground-state energies in MeV of the $^{4}$He nucleus obtained from the AV6P, AV8P, AV6P+UIX, and AV8P+UIX Hamiltonians with the AFDMC method. The VMC results correspond to using the AFDMC variational state with the linearized correlation operator of Eq.~\eqref{eq:corr}. The experimental ground-state energy is $-28.30$ MeV.\label{tab:energies_4He}}
\end{table}

\begin{figure}[t]
    \centering
    \includegraphics[width=\columnwidth]{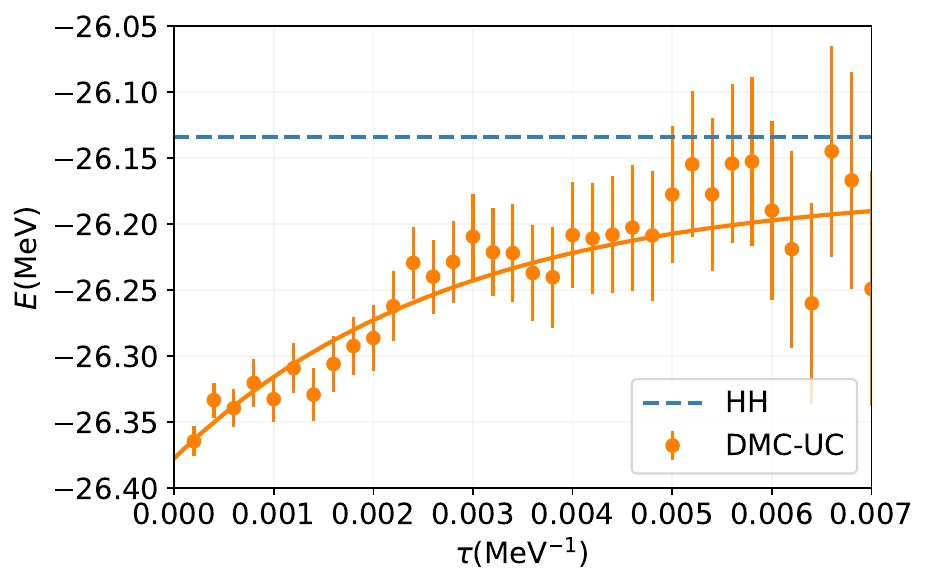}
    \caption{The orange solid circles represent the ADMC ground-state energy of $^4$He during the imaginary-time propagation. The input Hamiltonian only contains the AV6P potential. The solid orange line is the exponential fiting function. The HH reference value is displayed by the dashed blue line.}
    \label{fig:he4_trans_av6p}
\end{figure}

At first, we focus on the $^4$He nucleus, where we can carry out benchmark calculations between the AFDMC and HH methods. In the AV8P case, we estimate the uncertainty associated with using the replacement of Eq.\eqref{eq:spin_orbit_prop} to include the isospin-dependent spin-orbit term in the AFDMC imaginary-time propagator by varying the parameter $\alpha$ by 10\% around its best value ---  we find $\alpha=-2.825$ for AV8P and  $\alpha=-2.675$ AV8P+UIX cases, respectively. When the UIX potential is included in the Hamiltonian, we also vary $\beta$ by 10\% --- $\beta=1.583$ for both the AV6P+UIX and AV8P+UIX Hamiltonians. As a consequence, both the DMC-CP and DMC-UC errors for the AV8P, AV6P+UIX, and AV8P+UIX Hamiltonians are appreciably larger than for AV6P alone. Performing the unconstrained propagation improves the agreement between the AFDMC and HH methods, which is most apparent in the AV6P case, as it can be {\it exactly} included in the imaginary-time propagator. To better appreciate the bias introduced by the CP approximation, in Figure~\ref{fig:he4_trans_av6p}, we display the value of the ground-state energy of $^4$He during the unconstrained imaginary time propagation for the AV6P force. After about $0.007$ MeV$^{-1}$, the DMC-UC ground-state energy agrees with the highly accurate HH value. We note that in contrast with the fixed node approximation employed in condensed-matter applications~\cite{foulkes:2001}, the constrained propagation breaks the variational principle~\cite{Lynn:2017fxg}. The value of the fit and its error are reported in Table \ref{tab:energies_4He}. For the Hamiltonians that include spin-orbit NN interactions, the DMC-UC propagation produces energies about $0.5$ MeV less bound than the HH ones. However, the uncertainty associated with approximating the imaginary-time propagator as in Eq.~\eqref{eq:spin_orbit_prop} dominates this difference.

As already noted in Ref.\cite{Wiringa:2002ja}, both AV6P and AV8P alone underbind $^4$He by about $2$ MeV and $3$ MeV, respectively. In $^4$He, the attractive $V_{ijk}^\Delta$ contribution of UIX is significantly more important than the repulsive $V_{ijk}^R$, so overall UIX overbinds this nucleus by about $3$ MeV and $2$ MeV when used in combination with the AV6P and AV8P potentials, respectively. These findings are consistent with the fact that the full AV18 interaction is less attractive than both AV8P and AV6P in $^4$He\cite{Wiringa:2002ja}. 

\begin{table}[b]
\renewcommand{\arraystretch}{1.25} 
\centering
\begin{tabular}{c | c c c c }
\hline
\hline
Hamiltonian &  VMC & DMC-CP &  DMC-UC\\
\hline
 AV6P & -51.9(1)  & -101.9(2) & - \\
 AV8P & -40.2(1)  & -96(1)  & - \\
 AV6P+UIX & -45.9(2) & -108(8) & -113(9)\\
 AV8P+UIX & -40.2(4) & -104(8) & -112(9) \\
\hline
\end{tabular}
\caption{Ground-state energies in MeV of the $^{16}$O nucleus obtained from the AV6P, AV8P, AV6P+UIX, and AV8P+UIX Hamiltonians with the AFDMC method. The experimental ground-state energy is $-127.62$ MeV.\label{tab:energies_16O}}
\end{table}

The ground-state energies of $^{16}$O obtained from the AV6P, AV8P, AV6P+UIX, and AV8P+UIX Hamiltonians are listed in Table~\ref{tab:energies_16O}. To account for isospin-dependent spin-orbit terms in the imaginary-time propagator, we take $\alpha=-2.975$ for AV8P alone and $\alpha=-2.805$ for AV8P+UIX. On the other hand, we find the optimal values $\beta$ for the AV6P+UIX and AV8P+UIX Hamiltonians to be $\beta = 1.600$ and $\beta=1.650$, respectively. By comparing the DMC-CP and VMC ground-state energies, it is clear how the imaginary-time diffusion significantly improves them, much more than for $^4$He. This improvement is mostly due to the fact that the linearized spin-isospin correlations of Eq.\eqref{eq:corr} violate the factorization theorem, and hence become less accurate as the number of nucleons increases. Due to their significant computational cost, we carry out DMC-UC calculations only for the Hamiltonians that include a 3N force, namely AV6P+UIX and AV8P+UIX. The unconstrained propagation lowers the energies by about $5$ MeV for both Hamiltonians, indicating again the need for more sophisticated variational wave functions than the ones used here. The sizable uncertainties in both the DMC-CP and DMC-UC results are dominated by the approximations made in the imaginary-time propagator to include spin-isospin dependent spin-orbit terms and cubic spin-isospin operators—see Eqs.\eqref{eq:spin_orbit_prop} and~\eqref{eq:cubic_prop}.

For both AV6P+UIX and AV8P+UIX, $^{16}$O turns out to be underbound compared to experiment. This finding is consistent with the results obtained within the cluster variational Monte Carlo method~\cite{Lonardoni:2017egu}, although the AV18+UIX Hamiltonian was employed in that work. It also aligns with the observation that AV18+UIX underbinds light nuclei with more than $A=4$ nucleons~\cite{Pieper:2008rui,Carlson:2014vla}, as well as with infinite nuclear matter results~\cite{Akmal:1998cf,Lovato:2010ef}. Simply refitting the parameters $A_{2\pi}$ and $A_R$ is unlikely to resolve the overbinding of $^4$He and underbinding in heavier systems. Following the approach outlined in Ref.~\cite{Gnech:2023prs}, reducing the range of $V_{ijk}^R$ while increasing the magnitude of $A_R$ is likely to improve the agreement with experiments. To this end, in future work, we plan to replace the function $T^2(r)$ with the shorter-range Wood-Saxon functions $W(r)$ that define the shape of the core of the AV18 interaction.

\subsection{Spatial distributions and radii}
Within the AFDMC, expectation values of operators that do not commute with the Hamiltonian, such as the spatial and momentum distributions, are estimated at first order in perturbation theory as 
\begin{equation}
\frac{\langle\Psi(\tau) | O | \Psi(\tau)\rangle}{\langle\Psi(\tau) | \Psi(\tau)\rangle} \simeq 
2 \frac{\langle\Psi_V | O | \Psi(\tau)\rangle}{\langle\Psi_V | \Psi(\tau)\rangle} - \frac{\langle\Psi_V | O | \Psi_T\rangle}{\langle\Psi_V | \Psi_V\rangle}\, .
\label{eq:pc}
\end{equation}
Hence, in addition to controlling the fermion-sign problem, the accuracy of the variational state plays an important role to evaluate observables that do not commute with the Hamiltonian. In the following, we will present results for the single-nucleon density, which is defined as
\begin{align}
	\rho_{N}(r) &=\frac{1}{4\pi r^2}\langle \Psi_0 \big|\sum_i \mathcal \delta(r-|\mathbf{r}_i|)P_p(i)|\Psi_0 \rangle\,.
	\label{eq:rho_N}
\end{align}
Center of mass contaminations are automatically removed by $\mathbf{r}_i \to \mathbf{r}_i -\mathbf{R}_{\rm CM}$, where $\mathbf{R}_{\rm CM} = \sum_{i}\mathbf{r}_i/A$ is the center of mass coordinate of the nucleus~\cite{Massella:2018xdj}. In the above equation, we introduced the proton and neutron projector operators
\begin{equation}
P_p(i) = \frac{1 + \tau_i^z}{2}\quad,\quad P_n(i) = \frac{1 - \tau_i^z}{2}
\end{equation}

\begin{figure}[!t]
    \centering
    \includegraphics[width=\columnwidth]{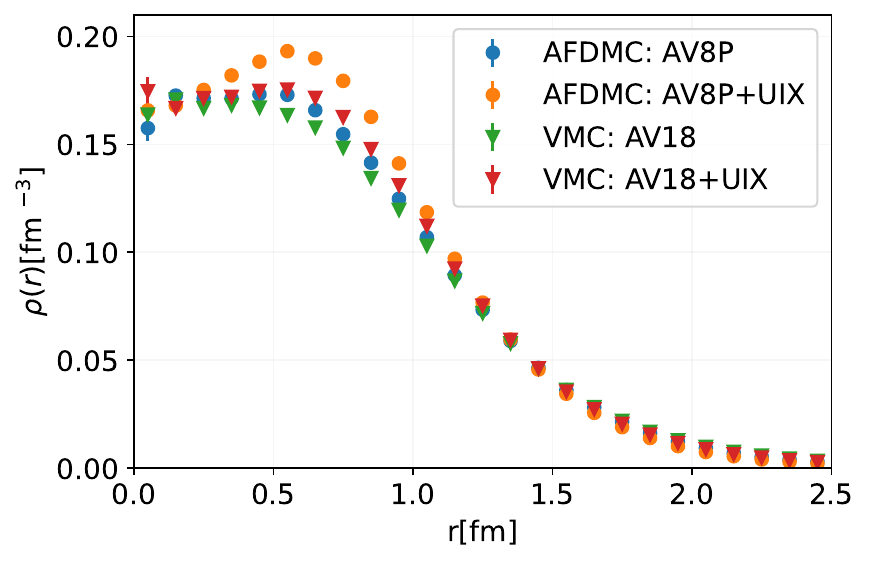}
    \caption{Point nucleon density of $^4$He as obtained with the AFDMC using the AV8P and AV8P+UIX interactions compared with the VMC results with the AV18 and AV18+UIX Hamiltonians.}
    \label{fig:he4_density}
\end{figure}

Figure~\ref{fig:he4_density} illustrates the AFDMC point-proton density of $^4$He as obtained from the AV8P and AV8P+UIX Hamiltonians. We compare the AFDMC results with VMC calculations carried out with the AV18 and AV18+UIX Hamiltonians~\cite{Pieper:2001ap,Pieper:2001mp}. When NN forces only are included, we observe excellent agreement between the VMC and the AFDMC. On the other hand, the effect of the 3N potential appears to be stronger in AFDMC than in the VMC, as it enhances the point-proton density at around $r=0.5$ fm. The corresponding charge radii, listed in Table~\ref{tab:he4_radii}, are consistent with this behavior. The expectation value of the charge radius is derived from the point-proton radius using the relation~\cite{Lonardoni:2017egu,Lonardoni:2018nob}
\begin{align}
\langle r_{\rm ch}^2\rangle=\langle r_{\rm pt}^2\rangle + R_{\rm p}^2 + \frac{A-Z}{Z} R_{\rm n}^2 + \frac{3\hbar^2}{4M_p^2 c^2}
\end{align}
where $r_{\rm pt}$ is the calculated point-proton radius, $R_{\rm p}^2
= 0.770(9)$ fm$^2$ the proton radius~\cite{ParticleDataGroup:2012pjm}, $R_{\rm n}^2 = -0.116(2)$ fm$^2$ the neutron radius~\cite{ParticleDataGroup:2012pjm}, and $3\hbar^2 / (4M_p^2 c^2) = 0.033$ fm$^2$ the Darwin-Foldy correction~\cite{Friar:1997js}.

Note that we neglect both the spin-orbit correction of Ref.\cite{Ong:2010gf} and two-body terms in the charge operator\cite{Lovato:2013cua}.

Consistent with the ground-state energies of Table~\ref{tab:energies_4He}, the theoretical uncertainty of the AFDMC calculations accounts for the approximations made in the imaginary-time propagator.

\begin{table}[!b]
\renewcommand{\arraystretch}{1.25} 
\centering
\begin{tabular}{c c c c c c }
\hline
\hline
AV8P     & AV8P+UIX & AV18 & AV18+UIX & Exp.\\
\hline
1.69(4) & 1.64(6) & 1.734(3)   & 1.665(3)      & 1.676(3)   \\
\hline
\end{tabular}
\caption{Charge radii (in fm) of $^{4}$He obtained from the AV8P and AV8P+UIX Hamiltonians with the AFDMC method compared with the VMC results for AV18 and AV18+UIX Hamiltonian~\cite{Pieper:2001ap,Pieper:2001mp}.\label{tab:he4_radii}}
\end{table}

\begin{figure}[t]
    \centering
    \includegraphics[width=\columnwidth]{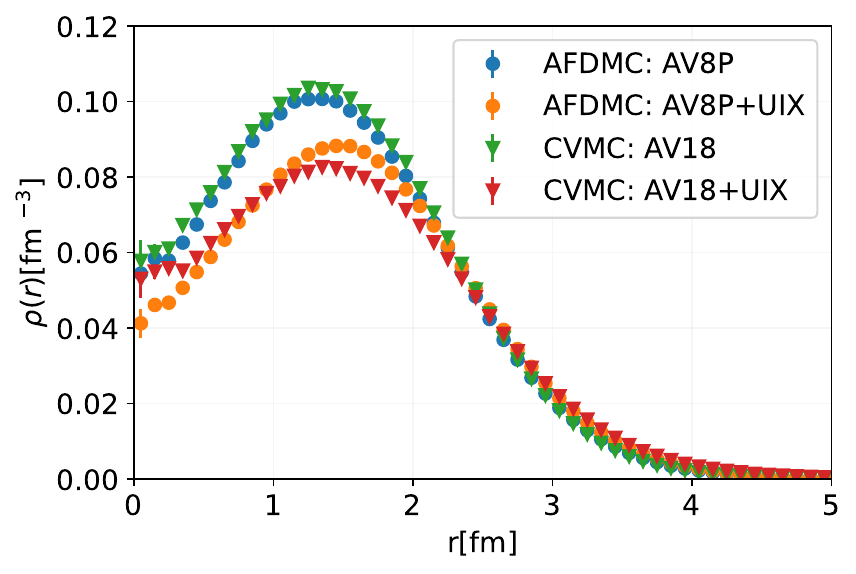}
    \caption{Point nucleon density of $^{16}$O as obtained with the AFDMC using the AV8P and AV8P+UIX interactions compared with the CVMC results with the AV18 and AV18+UIX Hamiltonians.}
    \label{fig:o16_density}
\end{figure}

The AFDMC point-proton density of $^{16}$O computed using the AV8P and AV8P+UIX Hamiltonians is displayed in Figure~\ref{fig:o16_density}. In this case, we compare the AFDMC results against the cluster variational Monte Carlo (CVMC) calculations carried out in Ref.\cite{Lonardoni:2017egu}. CVMC is a variational Monte Carlo method capable of computing nuclei with up to $A=40$ nucleons by utilizing a cluster expansion scheme for the spin-isospin dependent correlations present in the variational wave functions, considering up to five-body cluster terms. Thanks to this cluster expansion, there is no need to linearize the spin-dependent correlations as in Eq.\eqref{eq:corr}. Hence, the CVMC variational wave functions are more accurate than those employed as a starting point for AFDMC calculations. 

\begin{table}[b]
\renewcommand{\arraystretch}{1.25} 
\centering
\begin{tabular}{c c c c c c }
\hline
\hline
AV8P     & AV8P+UIX & AV18 & AV18+UIX & Exp.\\
\hline
2.58(6) & 2.61(9) & 2.538(2)   & 2.745(2)      & 2.699(5)   \\
\hline
\end{tabular}
\caption{Charge radii (in fm) of $^{16}$O obtained from the AV8P and AV8P+UIX Hamiltonians with the AFDMC method compared with the CVMC results for AV18 and AV18+UIX Hamiltonian~\cite{Lonardoni:2017egu}.\label{tab:o16_radii}}
\end{table}

There is excellent agreement between AFDMC and CVMC results, both for Hamiltonians including NN only and for the full NN+3N case. However, it should be noted that the repulsion brought about by the 3N potential appears to be stronger in AFDMC than in CVMC, resulting in a more prominent depletion of the density at short distances. These differences are likely attributed to the fact that in the CVMC method, three-body correlations in the variational wave function are treated at first-order in perturbation theory. On the other hand, although the same approximation applies to the AFDMC variational state, in AFDMC, correlations induced by the 3N potential are accounted for by the imaginary-time propagator—albeit the commutator term of UIX is approximated as discussed in Eq.\eqref{eq:cubic_prop}. This behavior is reflected in Table\ref{tab:o16_radii}, which shows the charge radius of $^{16}$O for the same Hamiltonians as discussed above. The 3N force, while being overall attractive, increases the charge radius by only about $0.03$ fm, primarily because of the repulsive $V^{R}_{ijk}$ term. However, it should be noted that these differences are much smaller than the uncertainties arising from using an approximated imaginary-time propagator to account for isospin-dependent spin-orbit terms and cubic spin-isospin operators.

\subsection{Momentum distributions}
Momentum distributions reflect features of both long- and short-distance nuclear dynamics and can provide useful insight into nuclear reactions, including inclusive and semi-inclusive electron and neutrino-nucleus scattering. The probability of finding a nucleon with momentum $\mathbf{k}$ and isospin projection $\tau=p,n$ in the nuclear ground-state is proportional to the density
\begin{align}
n_{\tau}(k) &= \int d\mathbf{r}_1^\prime  d\mathbf{r}_1 \dots d\mathbf{r}_A \Psi_0^*(\mathbf{r}_1^\prime, \mathbf{r}_2, \dots \mathbf{r}_A) \nonumber\\
& \times e^{-i\mathbf{k}\cdot(\mathbf{r}_1 -\mathbf{r}_1^\prime)} P_\tau(i) \Psi_0(\mathbf{r}_1, \mathbf{r}_2, \dots \mathbf{r}_A)\,,
\end{align}
where $\Psi(\mathbf{r}_1, \mathbf{r}_2, \dots \mathbf{r}_A) = \langle R | \Psi_0\rangle$ is a vector in spin-isospin space. 

\begin{figure}[!htb]
    \centering
    \includegraphics[width=\columnwidth]{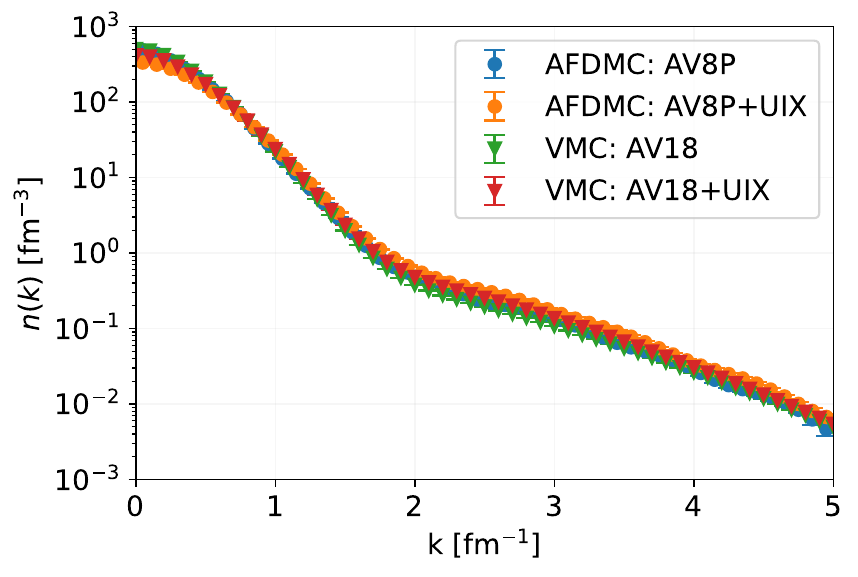}
    \caption{AFDMC proton momentum distribution of $^{4}$He for the AV8P and AV8P+UIX interactions compared with VMC results obtained with the AV18 and AV18+UIX Hamiltonians.}
    \label{fig:he4_momentum}
\end{figure}

Figure~\ref{fig:he4_momentum} shows the AFDMC proton momentum distribution of $^{4}$He as obtained using the AV8P and AV8P+UIX interactions, compared with VMC calculations that use the AV18 and AV18+UIX combinations of NN and 3N potentials. All the curves are close to each other, both in the low- and high-momentum regions, and they all exhibit a large tail extending to high momenta. The presence of this tail is a consequence of the high-resolution nature of the Argonne family of interactions, which generate high-momentum components in the wave function. Note that the similarity renormalization group has recently proven to quantitatively reproduce the high-resolution distributions using evolved operators and low-resolution wave functions~\cite{Tropiano:2024bmu}. The 3N potential reduces the momentum distribution at low momenta and further increases its tail. Overall, this effect appears to be more significant in the AFDMC than in the VMC, consistent with what is observed in the single-nucleon spatial density displayed in Figure~\ref{fig:he4_density}.  

\begin{figure}[b]
    \centering
    \includegraphics[width=\columnwidth]{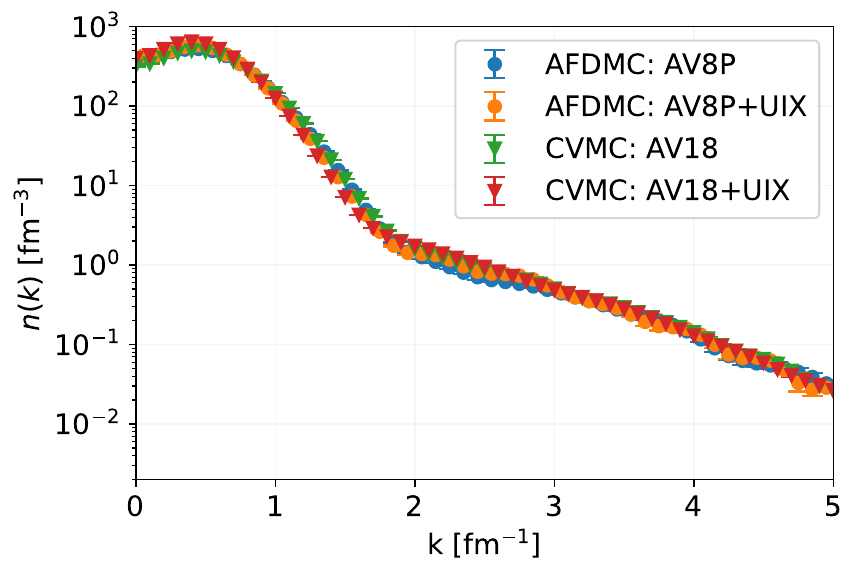}
    \caption{AFDMC proton momentum distribution of $^{16}$O for the AV8P and AV8P+UIX interactions compared with CVMC results obtained with the AV18 and AV18+UIX Hamiltonians~\cite{Lonardoni:2017egu}.}
    \label{fig:o16_momentum}
\end{figure}

Similar considerations, including the presence of a high-momentum tail, can be made for the proton momentum distribution of $^{16}$O, displayed in  Figure~\ref{fig:o16_momentum}. The AFDMC calculations agree well with the CVMC results from Ref.\cite{Lonardoni:2017egu}, largely independent of the input Hamiltonian. The accuracy of the CVMC method in computing the momentum distribution is particularly high, given the rapid convergence of the cluster expansions. This behavior is consistent with the recent similarity renormalization group analysis from Ref.~\cite{Tropiano:2024bmu}, where keeping only two-body terms in the unitary transformation operator reproduces VMC results at the percent level.

\subsection{Euclidean density responses}
In this work, we focus on the isoscalar density transition operator, defined as 
\begin{equation}
\rho(\mathbf{q}) = \sum_{j=1}^A e^{i \mathbf{q} \cdot \mathbf{r}_j}\, .
\label{eq:density_transition}
\end{equation}
Besides the obvious exception of the missing proton projection operator, the isoscalar density transition operator bears close similarities to the time component of the one-body electromagnetic current operator, which defines the electromagnetic longitudinal response of the nucleus. The latter quantity is critical for determining inclusive electron- and neutrino-nucleus scattering cross sections~\cite{Carlson:2001mp}, and to identify potential explicit quark and gluons effects in nuclear structure~\cite{Jourdan:1996np}. 

The corresponding response function is defined as 
\begin{equation}
R(\mathbf{q}, \omega) = \sum_f \langle |\Psi_f | \rho(\mathbf{q}) |\Psi_0\rangle|^2 \delta(\omega- E_f + E_0)\,.
\label{eq:response_def}
\end{equation}
where $\mathbf{q}$ and $\omega$ denote the momentum and energy transfer, respectively. In the above equation, $|0\rangle$ and $|f\rangle$ are the initial and final nuclear states with energies $E_0$ and $E_f$, respectively. In order to avoid computing all transitions induced by the current operator --- which is impractical except for very light nuclear systems~\cite{Shen:2012xz,Golak:2018qya} --- similar to previous Green's function Monte Carlo calculations, we infer properties of the response function from its Laplace transform~\cite{Carlson:1992ga,Carlson:2001mp}
\begin{equation}
E({\bf q},\tau) = \int_0^\infty d\omega\, e^{-\omega\tau} R({\bf q},\omega)\, .
\label{eq:euc_def}
\end{equation}
Leveraging the completeness of the final states of the reaction, the Euclidean response can be expressed as a ground-state expectation value:
\begin{equation}
E({\bf q},\tau)=\langle \Psi_0 | \rho^\dagger ({\bf q}) e^{-(H-E_0)\tau} \rho({\bf q}) | \Psi_0\rangle,
\end{equation}
where $H$ is the nuclear Hamiltonian.

Within the AFDMC, we evaluate the above expectation value employing the same imaginary-time propagation we use for finding the ground-state of the system. Since the isoscalar density transition operator defined in Eq.~\eqref{eq:density_transition} does not encompass spin-isospin operators, we can readily express the Euclidean response as 
\begin{equation}
E({\bf q},\tau) = \sum_{j,k=1}^A \langle \Psi_0 | e^{i \mathbf{q} \cdot [\mathbf{r}_j(\tau=0) - \mathbf{r}_k(\tau)] }| \Psi_0\rangle\, .
\end{equation}
where $\mathbf{r}_j(\tau)$ denotes the spatial coordinate of the $j$-th nucleon at imaginary time $\tau$. 

\begin{figure}[!htb]
    \centering
    \includegraphics[width=\columnwidth]{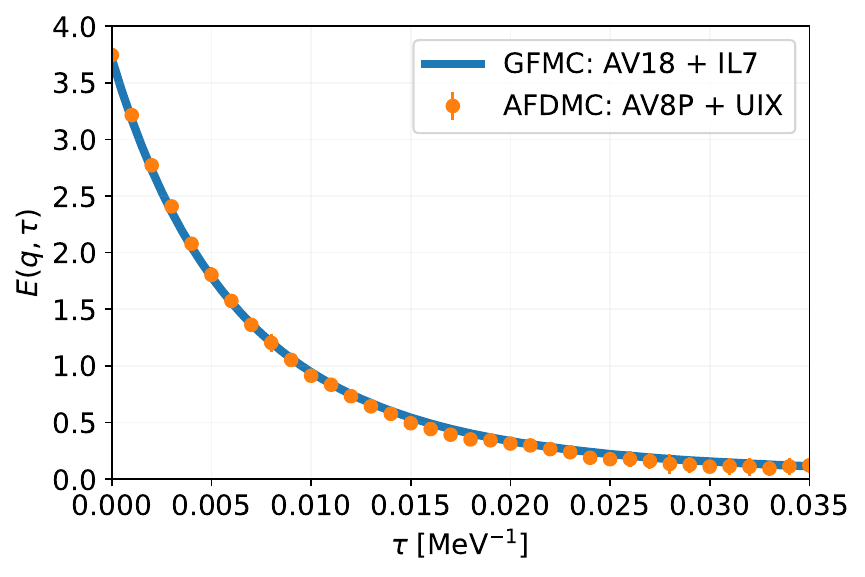}
    \caption{Euclidean isoscalar density response function of $^4$He at $q=600$ MeV obtained with the AFDMC (solid orange circles) and GFMC (blue solid band) methods using the AV8P+UIX and AV18+IL7 Hamiltonians, respectively. \label{fig:he4_euclidean_600} }
\end{figure}

In this work, as commonly done in previous Green's function Monte Carlo applications~\cite{Lovato:2013cua,Lovato:2020kba}, we approximate the ground-state $\Psi_0$ with the variational wave function $\Psi_V$. While this procedure is accurate for $^4$He, its applicability to $^{16}$O is more questionable, owing to the limitations inherent to using linearized spin-isospin dependent correlations. In future works, we plan on correcting these variational estimates within perturbation theory, as in Eq.~\ref{eq:pc}. For this reason, here we do not attempt inverting the Laplace transform either using Maximum Entropy~\cite{Bryan:1990, Jarrell:1996} or the recently developed machine-learning inversion protocols~\cite{Raghavan:2020bze,Raghavan:2023pav}. Rather, we focus on the computational feasibility of this quantity within the AFDMC, which can tackle larger nuclei than the GFMC, previously applied only up to $^{12}$C.  

In Figure~\ref{fig:he4_euclidean_600}, we compare the $^4$He Euclidean isoscalar density response function computed with the AFDMC and GFMC methods using as input the AV8P+UIX and AV18+IL7 Hamiltonians, respectively. Despite the different NN and 3N forces, the Euclidean response functions are close to each other up to the largest value of imaginary time we consider, $\tau = 0.5$ MeV$^{-1}$. This agreement is reassuring, as it corroborates both the AFDMC implementation of the Euclidean response-function calculation and the accuracy of the method, despite somewhat different input Hamiltonians. 

\begin{figure}[t]
    \centering
    \includegraphics[width=\columnwidth]{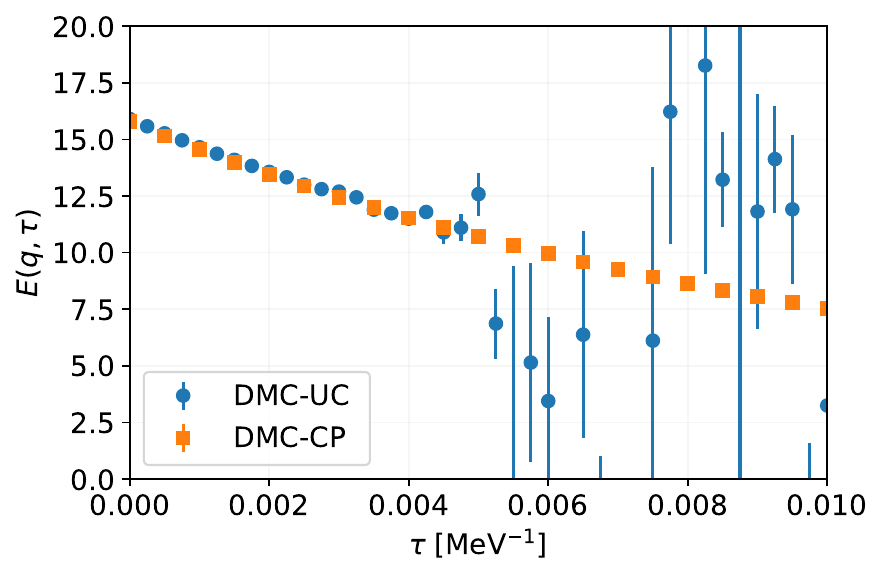}
    \caption{Euclidean isoscalar density response function of $^{16}$O at $q=400$ MeV with the AV8P+UIX Hamiltonian. The blue solid circles and the orange solid squares denote the unconstrained and constrained imaginary-time propagations, respectively.\label{fig:o16_euclidean_cp_uc}}
\end{figure}

The $^4$He Euclidean isoscalar density response function shown in Figure~\ref{fig:he4_euclidean_600} has been computed by carrying out DMC-UC propagations to remove the bias associated with the fermion sign problem. However, when tackling larger nuclei such as $^{16}$O, the unconstrained propagation yields sizable statistical fluctuations in the Euclidean response, primarily arising from the somewhat oversimplified correlation operator of Eq.\eqref{eq:corr}. In Figure~\ref{fig:o16_euclidean_cp_uc}, we display the Euclidean isoscalar density response function of $^{16}$He at $q=400$ MeV corresponding to the AV8P+UIX Hamiltonian. The DMC-UC results, denoted by the blue solid circles, start exhibiting large statistical fluctuations after around $\tau = 0.004$ MeV$^{-1}$. On the other hand, employing the same DMC-CP approximation used in ground-state calculations dramatically reduces the noise, making it possible to reach much larger imaginary times. In the region $\tau < 0.004$ MeV$^{-1}$, the DMC-CP and DMC-UC responses are consistent within statistical error, suggesting that the constrained-path approximation does not significantly bias the Monte Carlo estimates. Note that carrying out the DMC-UC calculations required about ten million CPU hours on Intel-KNL to keep the statistical noise under control, while the DMC-CP calculations are significantly cheaper. In the future, we plan to explore the contour-deformation techniques developed in Ref.~\cite{Kanwar:2023otc} to reduce the statistical uncertainty plaguing the unconstrained results.

\begin{figure}[b]
    \centering
    \includegraphics[width=\columnwidth]{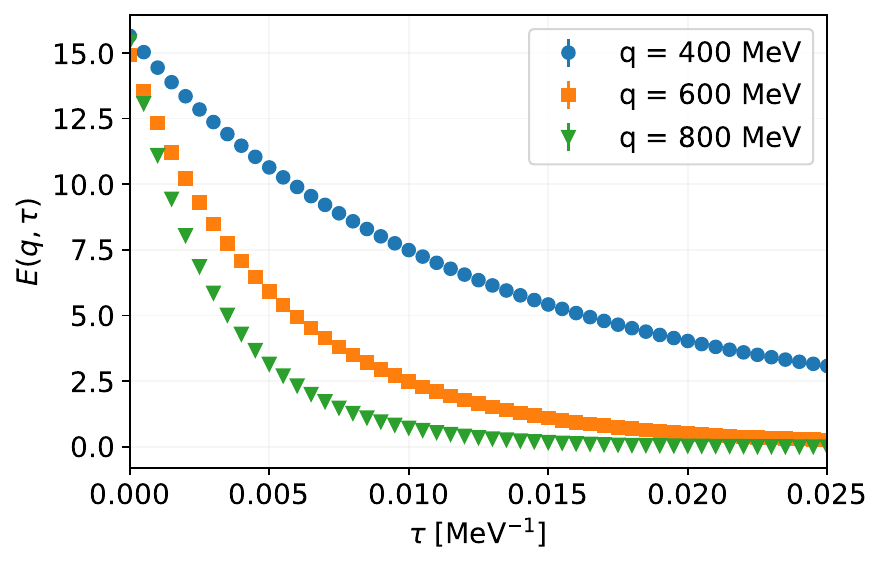}
    \caption{DMC-CP Euclidean isoscalar density response function of $^{16}$O at $q=400$ MeV (blue circles), $q=600$ MeV (orange squares), and  $q=800$ MeV (green triangles) computed with the AV8P+UIX Hamiltonian.}
    \label{fig:o16_euclidean_q.pdf}
\end{figure}

Figure~\ref{fig:o16_euclidean_q.pdf} displays the Euclidean isoscalar density response function at momentum transfers of $q=400$ MeV, $q=600$ MeV, and $q=800$ MeV, computed within the DMC-CP approximation using the AV8P+UIX Hamiltonian. Increasing the momentum transfer shifts the position of the quasi-elastic peak to higher energy transfers, thereby suppressing the Euclidean responses at large imaginary times. Additionally, the elastic contribution — corresponding to $\Psi_f = \Psi_0$ in Eq.~\eqref{eq:response_def} — is much smaller at $q=600$ MeV and $q=800$ MeV than at $q=400$ MeV, further contributing to the reduction of the Euclidean responses at large $\tau$. All these features are observed in the AFDMC curves, once again corroborating the accuracy of the novel methodology we developed.

\section{Conclusions}
\label{sec:conclusions}
We carried out AFDMC calculations of the ground-state static and dynamic properties of $^4$He and $^{16}$O using high-resolution phenomenological Hamiltonians from the Argonne family. For the $^4$He nucleus, we performed detailed benchmarks with the highly accurate hyperspherical harmonics method. This comparison corroborated the reliability of the AFDMC method, particularly the approximations made to treat isospin-dependent spin-orbit terms and cubic spin-isospin operators in the imaginary-time propagator. Nevertheless, these approximations contribute significantly to the uncertainties in our AFDMC results. To reduce them, we are currently working on a Restricted Boltzmann Machine representation of the imaginary-time propagator~\cite{Rrapaj:2020txq}, which is showing promising results and will be the subject of a future publication. Additionally, the AFDMC ground-state energies are affected by a non-negligible sign problem, particularly for $^{16}$O. To mitigate this issue, we plan to employ more accurate variational wave functions than those used here, leveraging neural-network quantum states~\cite{Adams:2020aax,Lovato:2022tjh}.

Our analysis confirms the inadequacy of both AV6P+UIX and AV8P+UIX Hamiltonians in describing simultaneously the ground-state properties of $^4$He and $^{16}$O nuclei. This finding is consistent with FHNC/SOC calculations of the energy per particle of isospin-symmetric nuclear matter at saturation density~\cite{Akmal:1998cf,Lovato:2010ef}, which indicate a binding energy about $3$ MeV lower than the empirical value. A potential solution to this problem, inspired by chiral effective field theory, involves replacing the squared Yukawa function in the repulsive 3N component with a shorter-ranged parameterization, consistent with the Wood-Saxon functions defining the core shape of the AV18 interaction. To this end, we plan to incorporate these shorter-ranged 3N forces in a simultaneous AFDMC/GFMC analysis of $^4$He, $^{12}$C, and $^{16}$O nuclei, alongside the neutron-matter equation of state. Our objective is to identify a high-resolution Hamiltonian capable of accurately describing nuclei up to $^{16}$O, as well as the high-density neutron-star matter found in the inner core of neutron stars. An alternative possibility is to include relativistic boost corrections in the Hamiltonian~\cite{Forest:1995sg}, which are known to change the strength of the 3N interaction in light nuclei~\cite{Forest:1995zz,Yang:2022esu} and might help reproducing saturation properties of isospin-symmetric nuclear matter~\cite{Akmal:1998cf,Yang:2024wsg}. 

We established a computational protocol for carrying out AFDMC calculations of the Euclidean isoscalar density response function of atomic nuclei, validating it against GFMC results of $^4$He. Treating $^{16}$O results in significantly higher statistical noise, which we controlled by employing the same constrained-path approximation used for computing the ground-state energy of the system. To achieve fully unbiased calculations, we intend to explore contour deformation techniques successfully applied to GFMC calculations of the isoscalar density response of the deuteron in Ref.~\cite{Kanwar:2023otc}. With the algorithmic foundations laid, we are now well-positioned to compute the electroweak response functions of $^{16}$O, including both one- and two-body current contributions, which are essential for evaluating inclusive electron and neutrino scattering cross-sections.

The calculations conducted in this work will provide important inputs to the ACHILLES neutrino event generator~\cite{Isaacson:2020wlx,Isaacson:2022cwh}. Specifically, we will supply ACHILLES with AFDMC samples of three-dimensional positions of protons and neutrons in the $^{16}$O nucleus. These samples fully encode correlation effects and, by definition, reproduce the one- and two-body isospin dependent spatial density distributions. The significance of including correlations in the propagation of the scattered particle throughout the nucleus will be examined by comparing against nuclear transparency data~\cite{Isaacson:2020wlx} and other relevant observables. Additionally, the momentum distributions computed in this study will serve as a normalization test for the single-nucleon spectral functions used in ACHILLES to compute the elementary vertex.

\section{Acknowledgements}
We thank R. B. Wiringa for providing us with the phase shifts of the Argonne potentials and for critically reading the manuscript. We are also grateful to R.~Schiavilla for the invaluable suggestions provided during the initial phase of this study. This work was supported by the U.S. Department of Energy (DOE), Office of Science, Office of Nuclear Physics under contract number DE-AC02-06CH11357 (A.L.) and by the SciDAC-NUCLEI (A.L., ) and SciDAC-NeuCol (A.L. and N.R.) projects. A.L. is also supported by DOE Early Career Research Program awards. A.G. acknowledges support from the DOE Topical Collaboration “Nuclear Theory for New Physics,” award No. DE-SC0023663 and Jefferson Lab that is supported by the U.S. Department of Energy, Office of Nuclear Science, under Contracts No. DE-AC05-06OR23177.  Quantum Monte Carlo calculations were performed on the parallel computers of the Laboratory Computing Resource Center, Argonne National Laboratory, the computers of the Argonne Leadership Computing Facility via ALCC and INCITE grants. Hyperspherical Harmonic calculations were performed at NERSC, a
U.S. Department of Energy Office of Science User Facility
located at Lawrence Berkeley National Laboratory, operated under Contract No. DE-AC02-05CH11231 using
NERSC award NP-ERCAP0023221.

\bibliography{biblio}

\end{document}